\documentclass[mathleft
]{an}
\usepackage{graphicx}
\usepackage{times}
\usepackage{natbib}
\bibpunct[, ]{(}{)}{;}{a}{}{,}
\begin{document}

\Pagespan{789}{}
\Yearpublication{2006}%
\Yearsubmission{2005}%
\Month{11}%
\Volume{999}%
\Issue{88}%

\title{Dynamics of the TrES-2 system}

\author{F. Freistetter\inst{1}\fnmsep\thanks{Corresponding author:
  \email{florian@astro.uni-jena.de}\newline}
\and  {\'A}. S\"uli\inst{2}
\and B. Funk\inst{3}
}
\titlerunning{Dynamics of the TrES-2 system}
\authorrunning{F. Freistetter et al.}
\institute{
Astrophysikalisches Institut und Universit\"atssternwarte Jena, Schillerg\"a{\ss}chen 2-3,
D-07745 Jena, Germany
\and 
Department of Astronomy E\"otv\"os Lor\'and University P\'azm\'any P\'eter s\'et\'any 1/A 1117 Budapest, Hungary
\and 
Institut f\"ur Astronomie, Universit\"at Wien, T\"urkenschanzstrasse 17, A-1180 Wien, Austria}

\received{2008}
\accepted{2008}
\publonline{later}


\abstract{The TrES-2 system harbors one planet which was discovered with the transit technique. In this work we investigate the dynamical behavior of possible additional, lower-mass planets. We identify the regions where such planets can move on stable orbits and show how they depend on the initial eccentricity and inclination. We find, that there are stable regions inside and outside the orbit of TrES-2b where additional, smaller planets can move. We also show that those planets can have a large orbital inclination which makes a detection with the transit technique very difficult.}

\maketitle

\section{Introduction}

The planet TrES-2b was discovered in 2006~\citep{Odonovan06}. It orbits a G0V star with approximately the same mass and radius as the Sun. The planet has a mass of 1.198 M$_{\mathrm{JUP}}$ and a radius of 1.22 R$_{\mathrm{JUP}}$. Its orbit is circular and has a semi-major axis of 0.0367 AU (see table~\ref{tab1} for all parameters and the corresponding error bars.)

\begin{table}
\caption{Basic data of the TrES-2 system from~\cite{Odonovan06, Winn08}}
\label{tab1}
\begin{tabular}{lr} 
\multicolumn{2}{c}{TrES-2b}\\ \hline 
Mass & $1.198 \pm 0.053 M_{\mathrm{JUP}}$  \\
Semimajor axis & $0.0367 \pm 0.0012$ AU\\
Orbital period & $2.470621 \pm 0.000017$ days\\
Eccentricity & $0$\\
Radius & $1.22 \pm 0.045 R_{\mathrm{JUP}}$\\
Time of transit & $ 2453957.63479 \pm 0.00038$ \\
Inclination & $83.9 \pm 0.22$ \\ \hline 
\end{tabular}
\end{table}

TrES-2b was discovered using the transit method where one tries to observe the decrease in luminosity of a star caused by a transiting planet. This method is very promising and current and future space-mission (e.g. {\tt CoRoT} and {\tt Kepler}) will use this method to largely increase the number of known transiting planets. Although it is possible to detect quite small planets via the transit technique the discovery of earth-like planets is still very difficult. The premises are better in systems were a transiting planet is already discovered. By investigating the variations in the transit time of a planet it is possible to identify also very small planets~\citep{Ford07, Ford06}. \\
It is thus desirable to have a detailed knowledge about the dynamical properties of the systems with known transiting planets. 
In this work, we investigate the gravitational influence of TrES-2b on a potential second planet in the system. Depending on the orbital elements of TrES-2b we can identify the zones of stability, where it could be possible to detect an additional planet.\\
Our work is motivated by the very fact that for the TrES-2 planet, transit time variations have been reported by~\cite{Odonovan06},~\cite{Raetz08a} and \cite{Raetz08b}.

\section{Numerical Setup}

The majority of planetary systems consist of a star and a giant planet orbiting in an eccentric orbit around it. TrES-2 is such a single-planet system and can be modeled by the elliptic restricted three-body problem, a relatively simple dynamical model, where two massive bodies, called the primaries, move in elliptic orbits
around their common center of mass, and a third body of negligible mass moves under their gravitational influence (for details see \cite{Szebehely1967}).
For the computation of the stability maps we varied both the orbital elements of the giant and the test particles and for more than 350 thousands initial conditions the dynamical character of the massless particle representing an earth-like planet was determined. To compute the stability maps, the method of the maximum eccentricity (ME) and the Lyapunov characteristic indicator (LCI) were used as main tools for stability investigations of massless terrestrial-like planets:

\begin{itemize}
\item
To indicate stability, the ME method uses a straightforward check based on the eccentricity. This action-like variable shows the probability of orbital crossing and close encounter of two bodies and therefore its value provides information on the stability of orbits. This simple check was already used in several
stability investigations, and was found to be a powerful indicator of the stability character of orbits \citep{dvorak03b,suli05,Nagy2006}. 
\item
As a complementary tool, we computed also the LCI, a well-known chaos
indicator. The LCI is the finite time approximation of the largest Lyapunov exponent (LCE), which is described in detail in e.g.~\cite{froeschle84} and \cite{lohinger93a}.
\end{itemize}

The  mass of $\mu$ of the TrES-2b system is approximately 0.0022097 solar masses which can be rounded to 0.002. The time span of the integration was $10^5$ revolutions of the primaries. For the integration of the system we applied an efficient variable-time-step algorithm, known as Bulirsch-Stoer integration method. The most important feature of the Bulirsch-Stoer algorithm for N-body simulations is that it is capable of keeping an upper bound on the local errors introduced due to taking finite time-steps by adaptively reducing the step size when interactions between the particles increase in strength. The parameter $\epsilon$ which controls
the accuracy of the integration was determined by preliminary test runs and was set to $10^{-8}$.

The giant planet's orbital elements are: semi-major axis $a_1$, eccentricity $e_1$, argument of periastron $\omega_1$, and mean anomaly $M_1$. The semi-major axis of the giant is $a_1=0.0367$ AU, and its eccentricity was 0. Since the giant revolves in a circular orbit, $\omega_1$ is undefined and the mean anomaly alone determines the position of the giant along the orbit. $M_1$ was varied between $0^\circ$ and $180^\circ$ with $\Delta M_1=45^\circ$.

The semi-major axis of the test particles ranges from 0.014 to 0.147 AU for stability maps in the $a-i$ plane and up to 0.183 AU for stability maps in the $a-e$ plane. The inner border for the test particles is very close to the star. Although earth-mass planets could exist here, due to the large temperatures they would certainly not be earth-like planets. The step-size $\Delta a = 1.8\cdot10^{-4}$ AU, $\Delta e= 0.02$ and $\Delta i= 1^\circ$. The upper limit of the grid in $e$ is 0.5 and $50^\circ$ in $i$.

To investigate the resonances in detail we also applied the restricted three body problem. In this case the integration of the Newtonian equations of motion was undertaken with a different method. We used the so called {\em Lie integration} \citep{Hanslmeier84,Lichtenegger84}), which implements an automatic step-size and is, because of the recurrence of the Lie-terms, a very precise integration method. Again the integration time was $10^5$ revolutions of the primaries. For the calculations we took into account the resonances up to the 7:4 mean motion resonance for the inner and the outer region. For the motion in resonances the initial starting position is quite important, therefore we have chosen 8 starting positions for the fictitious planets in a certain resonance with the mean anomaly M ranging from $0^{\circ}$ to $315^{\circ}$ with $\Delta M = 45^\circ$.

\begin{figure*}
\begin{center}
\includegraphics[width=80mm]{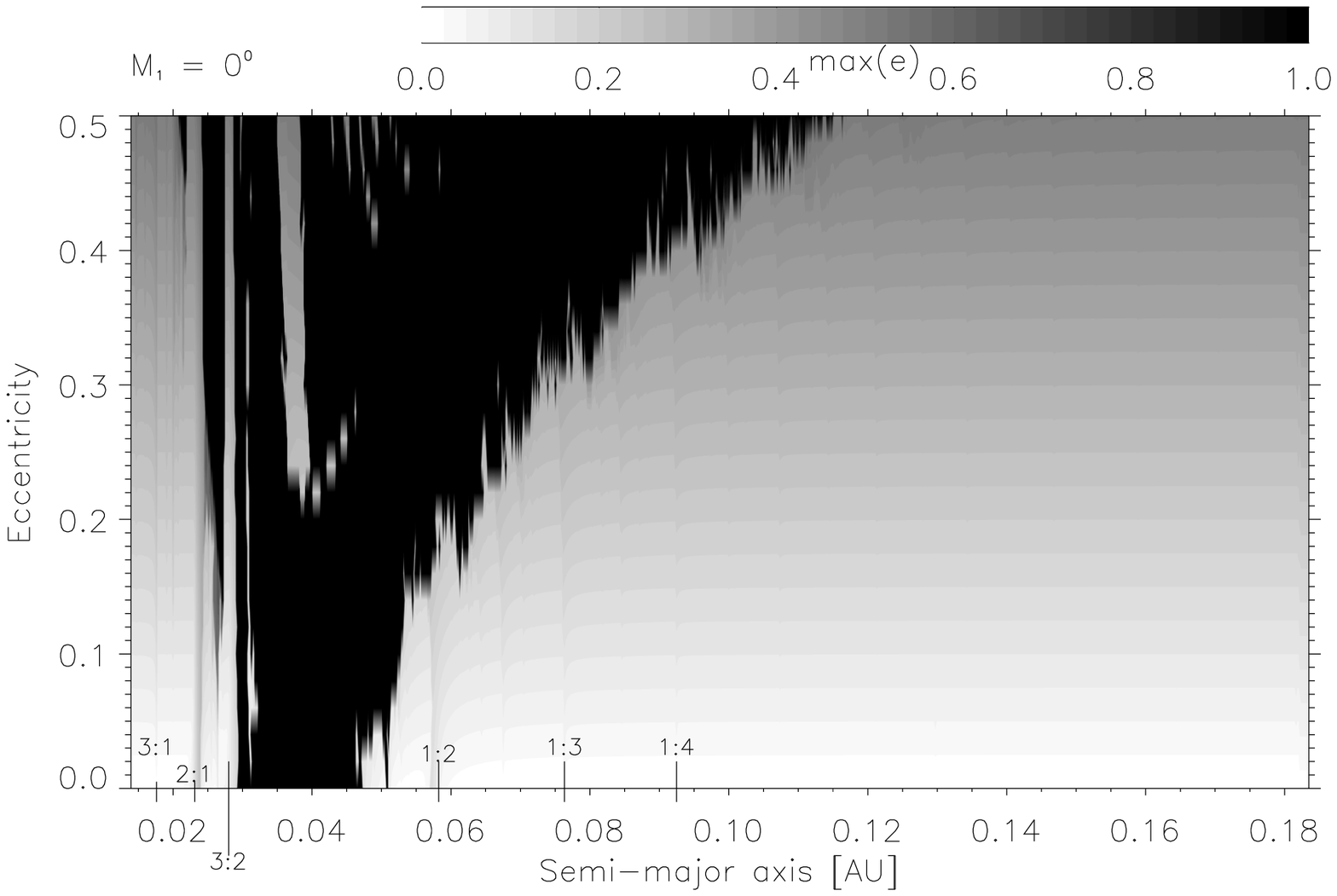}
\includegraphics[width=80mm]{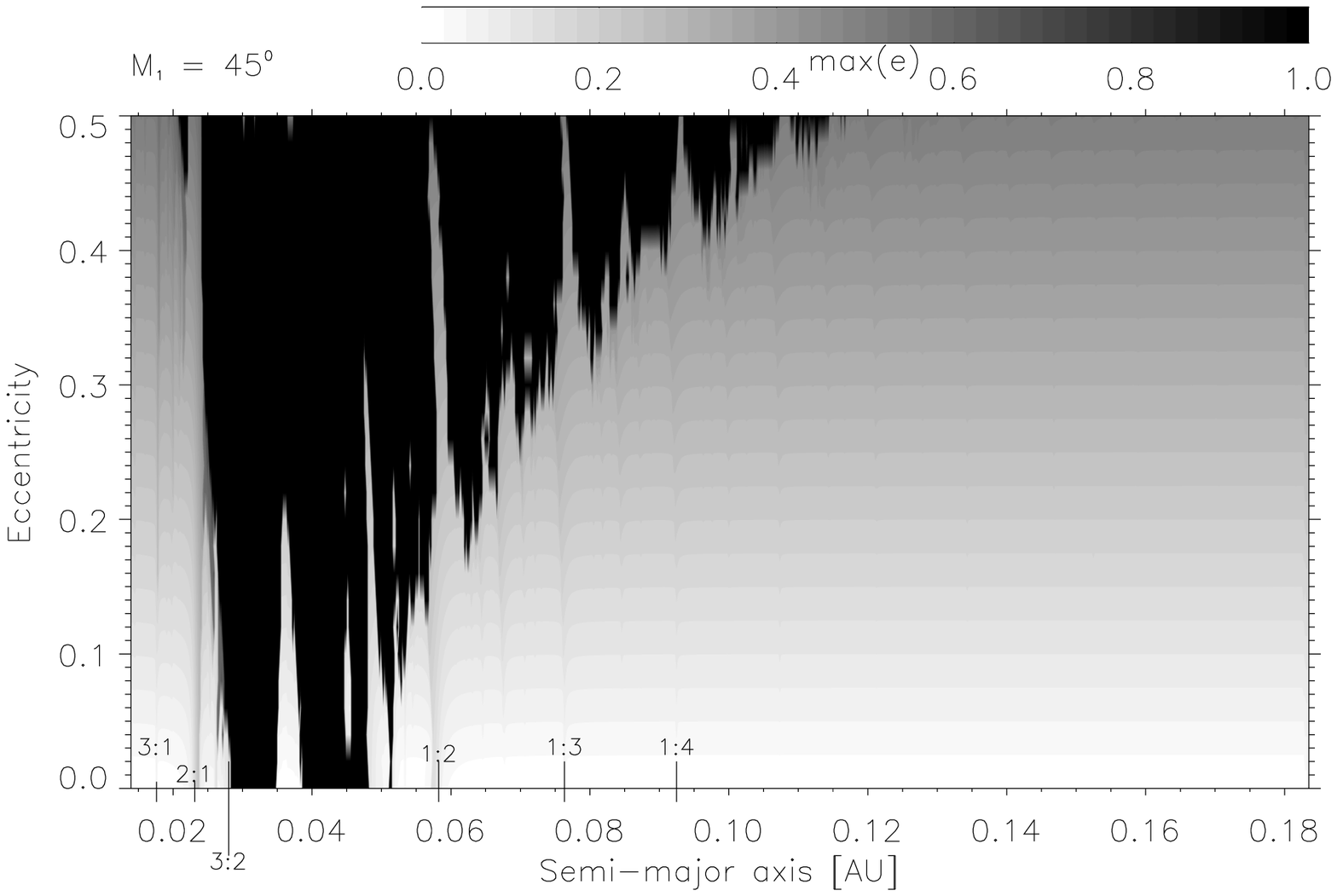}\\
\includegraphics[width=80mm]{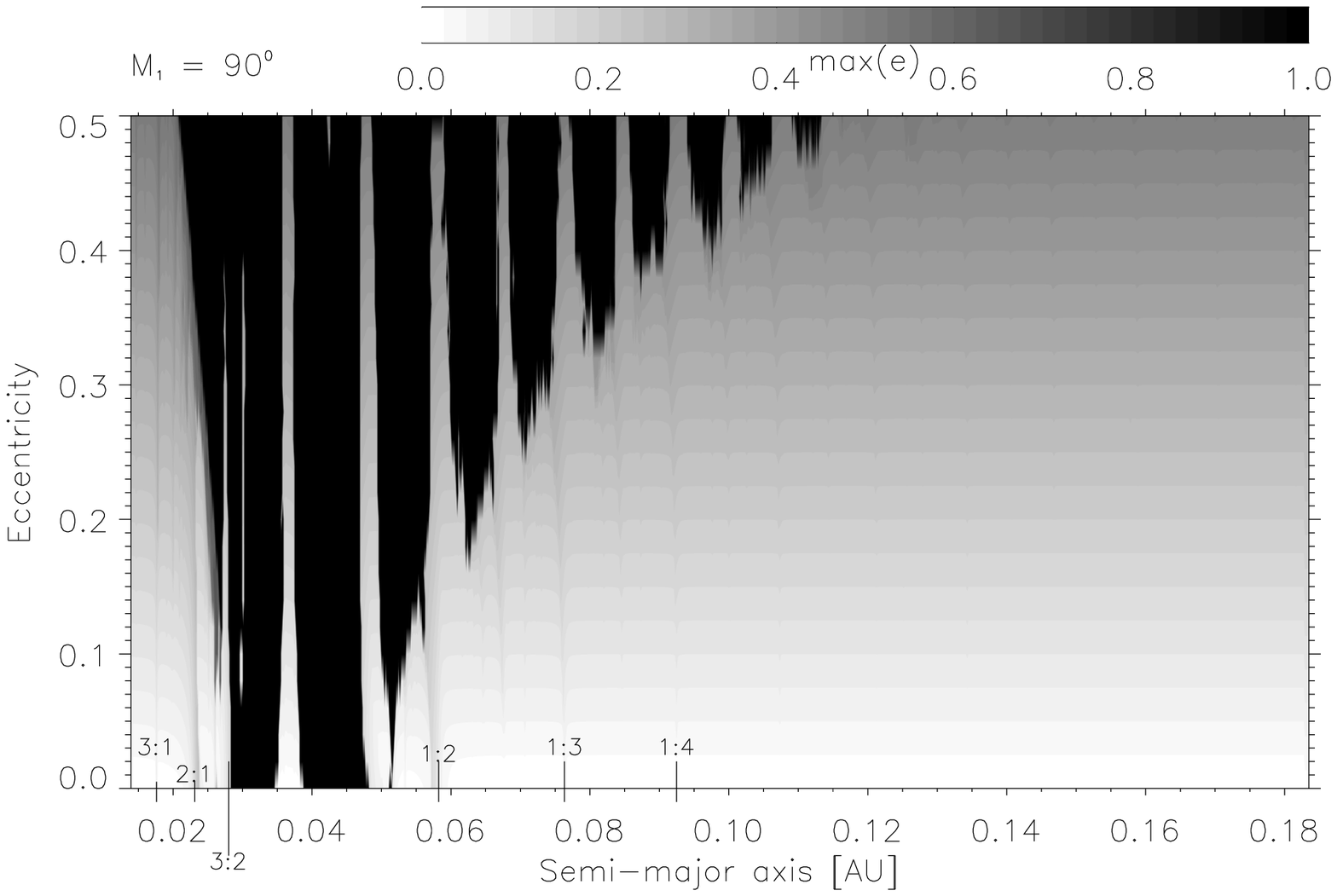}
\includegraphics[width=80mm]{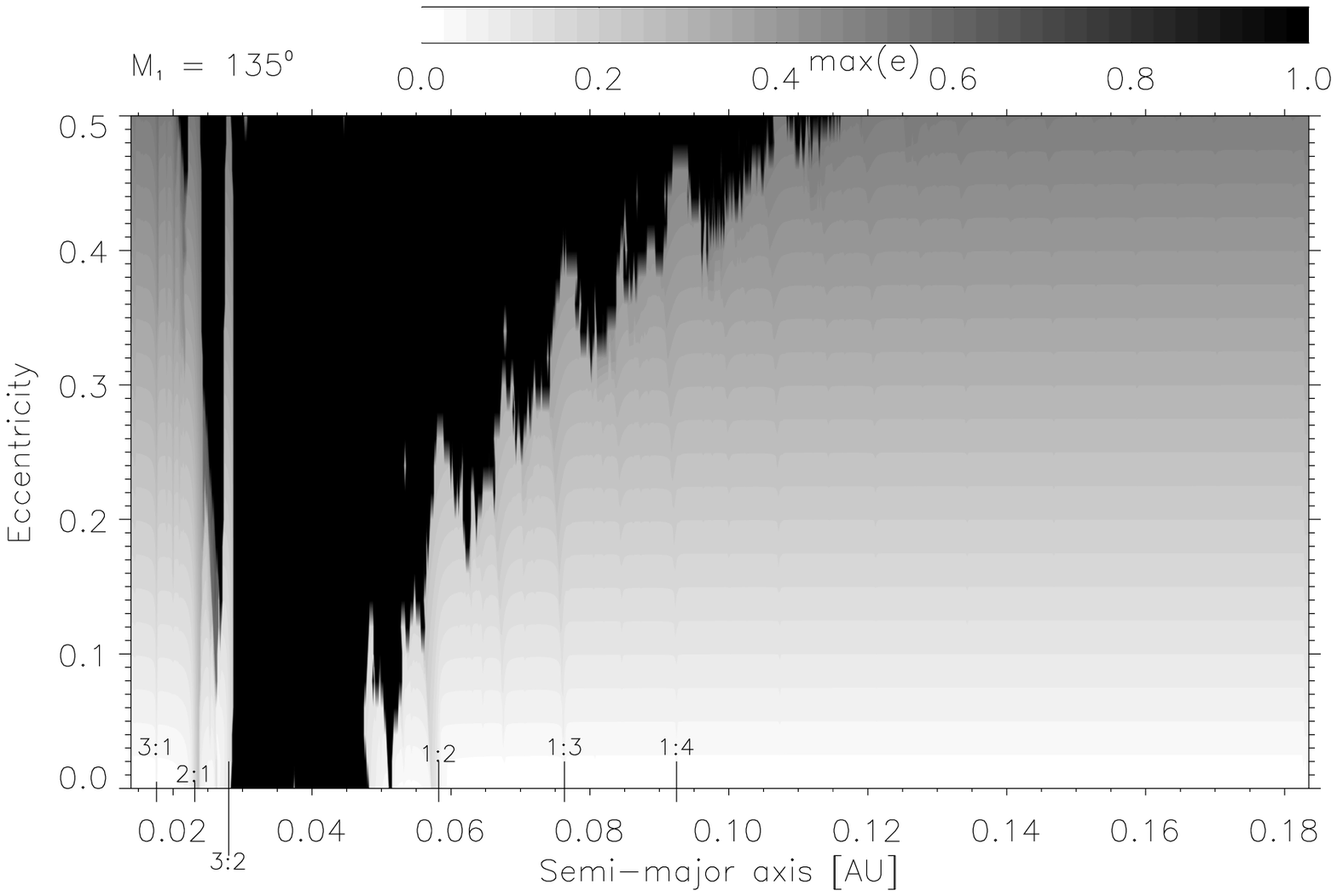}\\
\includegraphics[width=80mm]{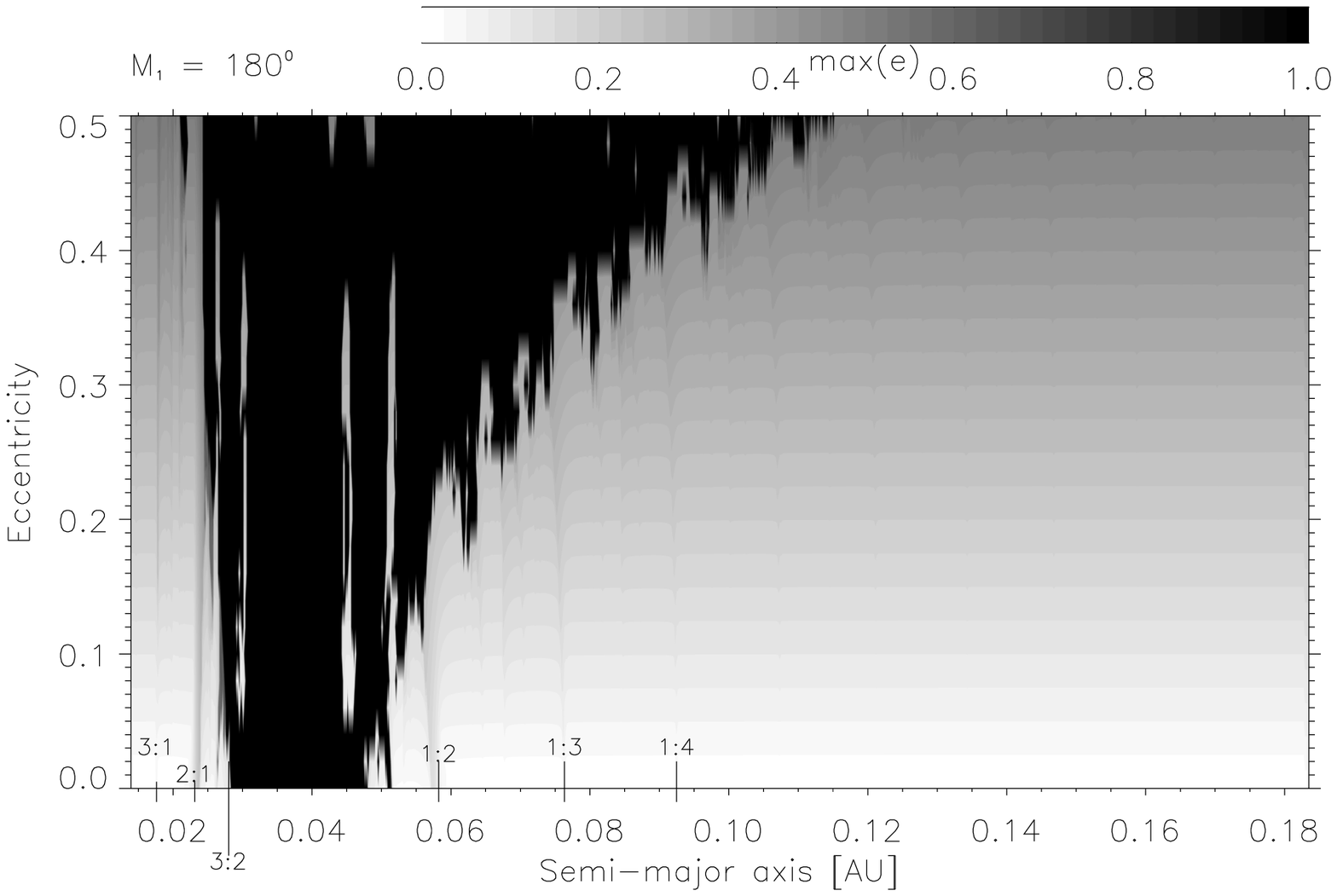}
\includegraphics[width=80mm]{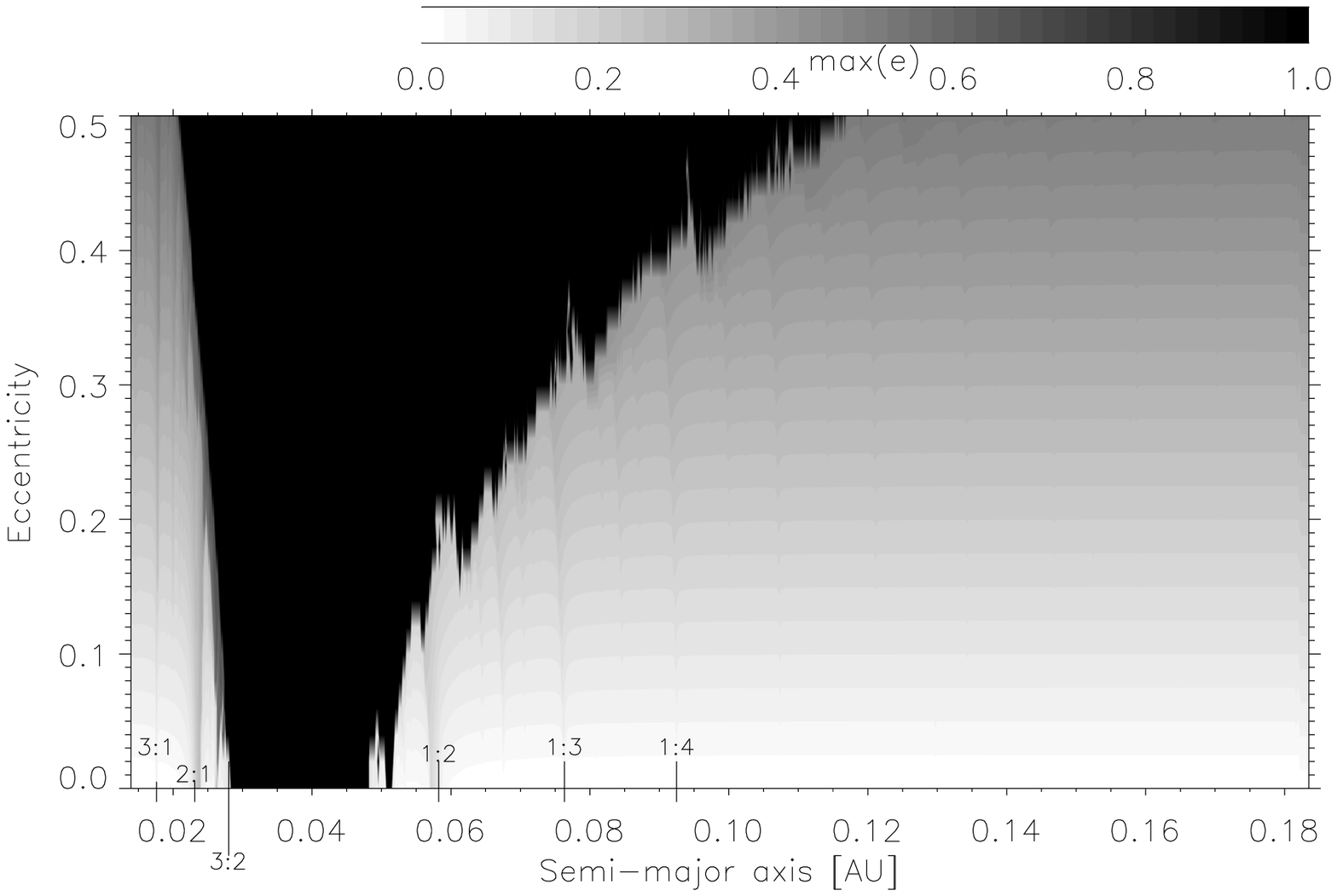}
\end{center}
\caption{Stability plots in the $a-e$ plane for different initial values of $M$ showing the maximum eccentricity (ME). From top left to bottom left: M=0$^\circ$, M=45$^\circ$, M=90$^\circ$, M=135$^\circ$, M=180$^\circ$. Bottom right: combination of all figures using the maximum ME over all values of M.}
\label{label1}
\end{figure*}

\section{Stability Regions}
\label{sec:stab}

Figure~\ref{label1} shows the stability plots for the $a-e$ plane. We investigated the small region between the star and the known planet and the region outside the orbit of TrES-2b up to semimajor-axes of 0.18 AU. The first 5 pictures show the stability region for different initial values of the mean ano\-maly of TrES-2b whereas for the last picture we took the global maximum eccentricity that was obtained for any value of $M_1$. \\
The overall shape of the stability is of course not influenced by the mean anomaly; it looks the same in all cases. One can see that the border of the inner region between the star and TrES-2b is almost independent of the initial eccentricity of the test particle. For initial semimajor axes smaller than 0.025 and initial eccentricities smaller than 0.25 the test particles stay stable on moderate eccentric orbits with $e < 0.25$. Particles with larger initial eccentricities can still stay on stable (eccentric) orbits further away from the planet ($a < 0.02$ AU). \\
The border of the stability region outside the orbit of TrES-2b shows a different behavior. For almost circular orbits the whole range between 0.05 an 0.18 AU shows stable motion. This region however is shrinking when the initial eccentricity is increased. The gravitational influence of TrES-2b reaches out to 0.12 AU; from this point on the stability region is independent from the initial eccentricity.

The influence of {\em mean-motion resonances} with TrES-2b is also visible in the figures. Depending on the initial value of $M_1$, one can observe additional stable regions corresponding to the position of the resonances. They are most stable for $M_1=90^\circ$; an initial value of 45$^\circ$ shows fewer stable resonances; for 0$^\circ$, 135$^\circ$ and 180$^\circ$ almost all resonances become unstable (for a detailed discussion of the stability of resonances see section~\ref{sec:res}).

\begin{figure*}
\begin{center}
\includegraphics[width=80mm]{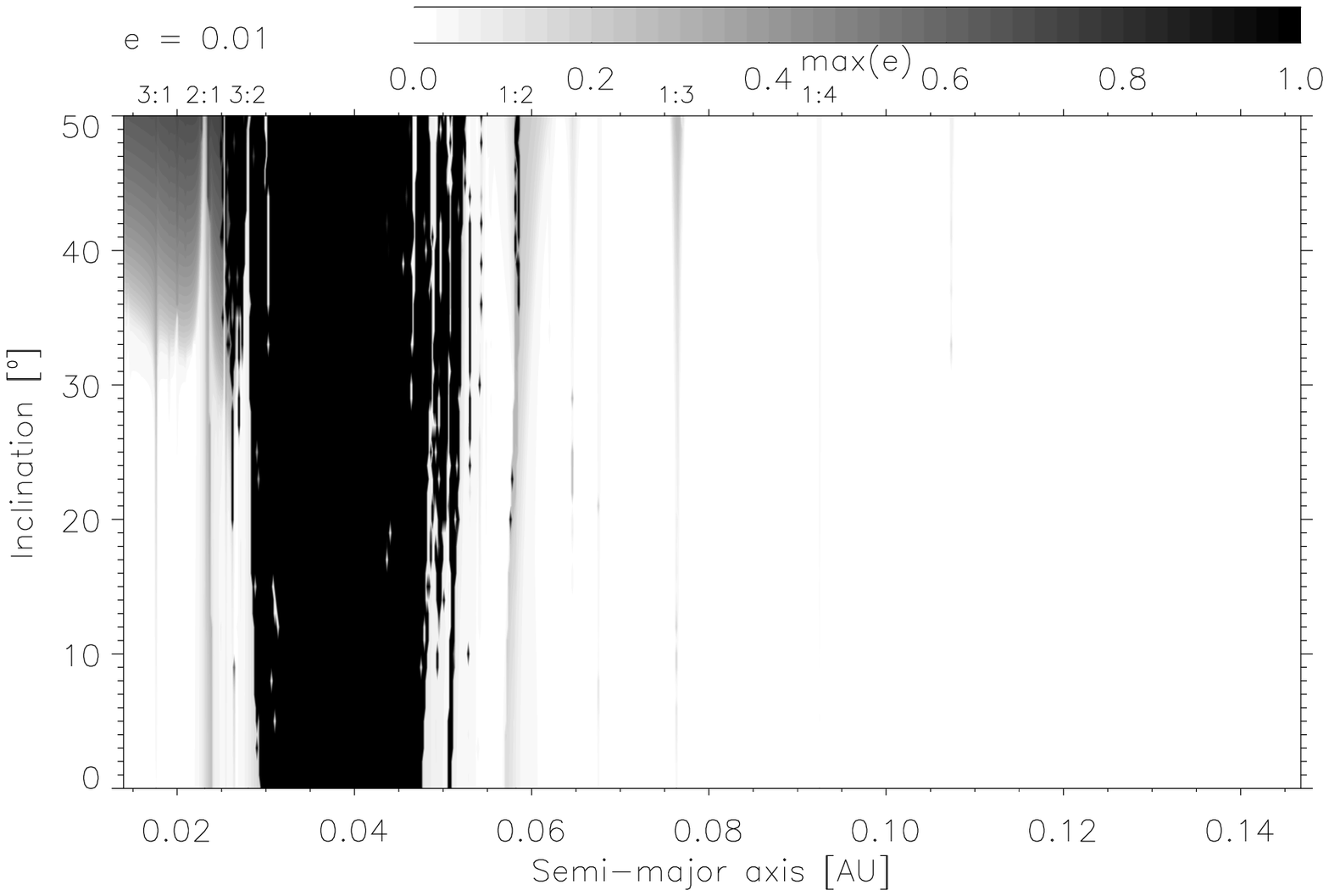}
\includegraphics[width=80mm]{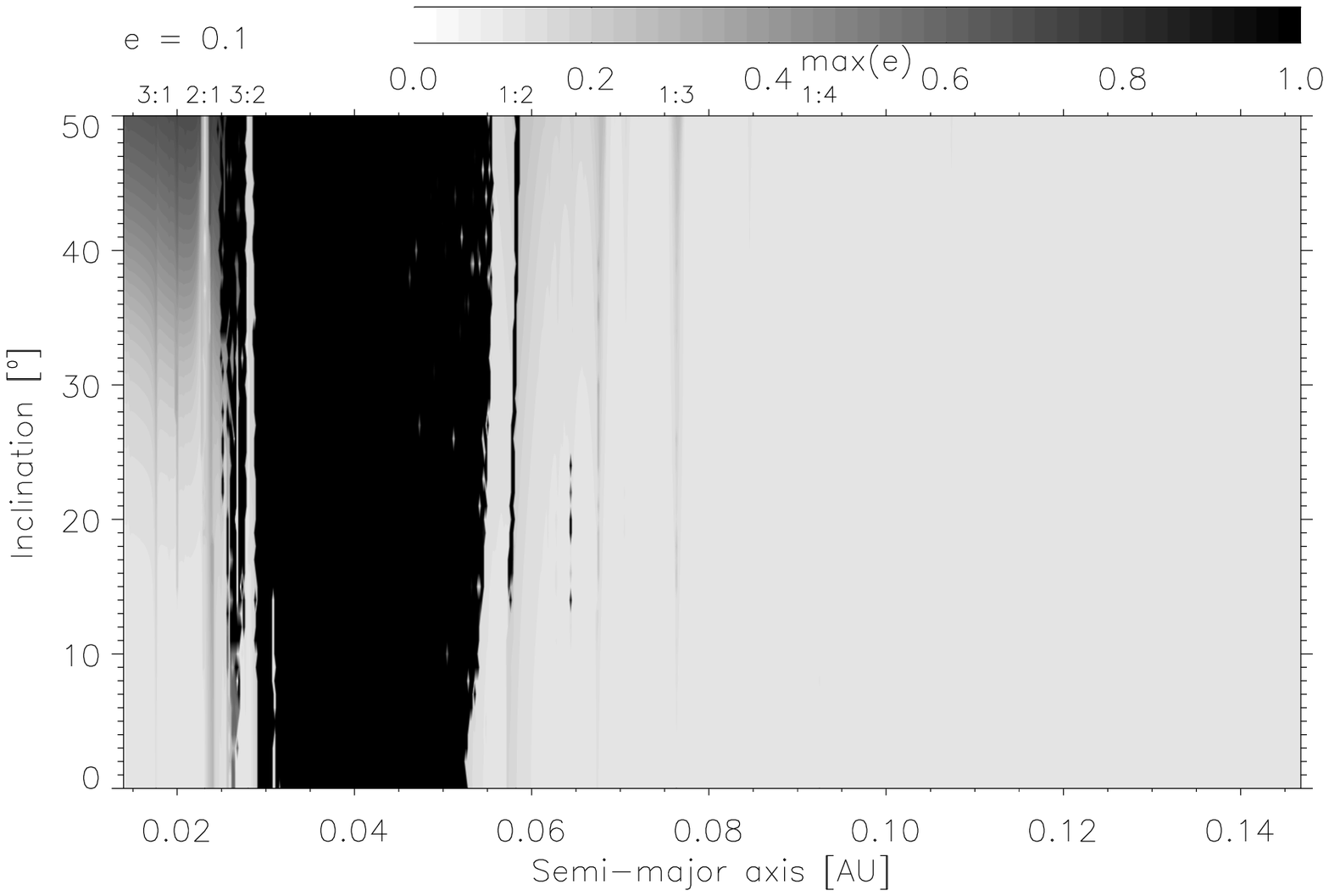}\\
\includegraphics[width=80mm]{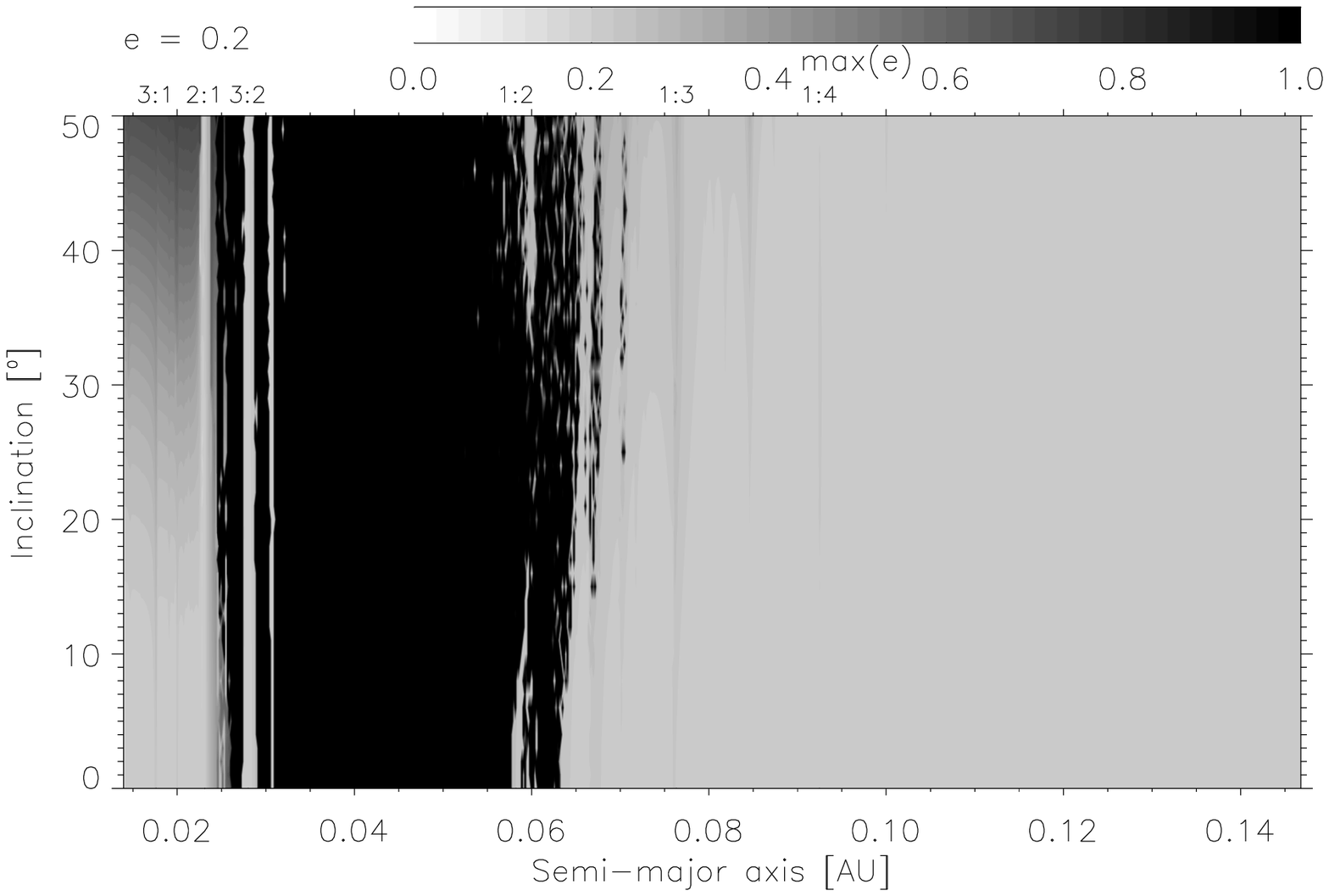}
\includegraphics[width=80mm]{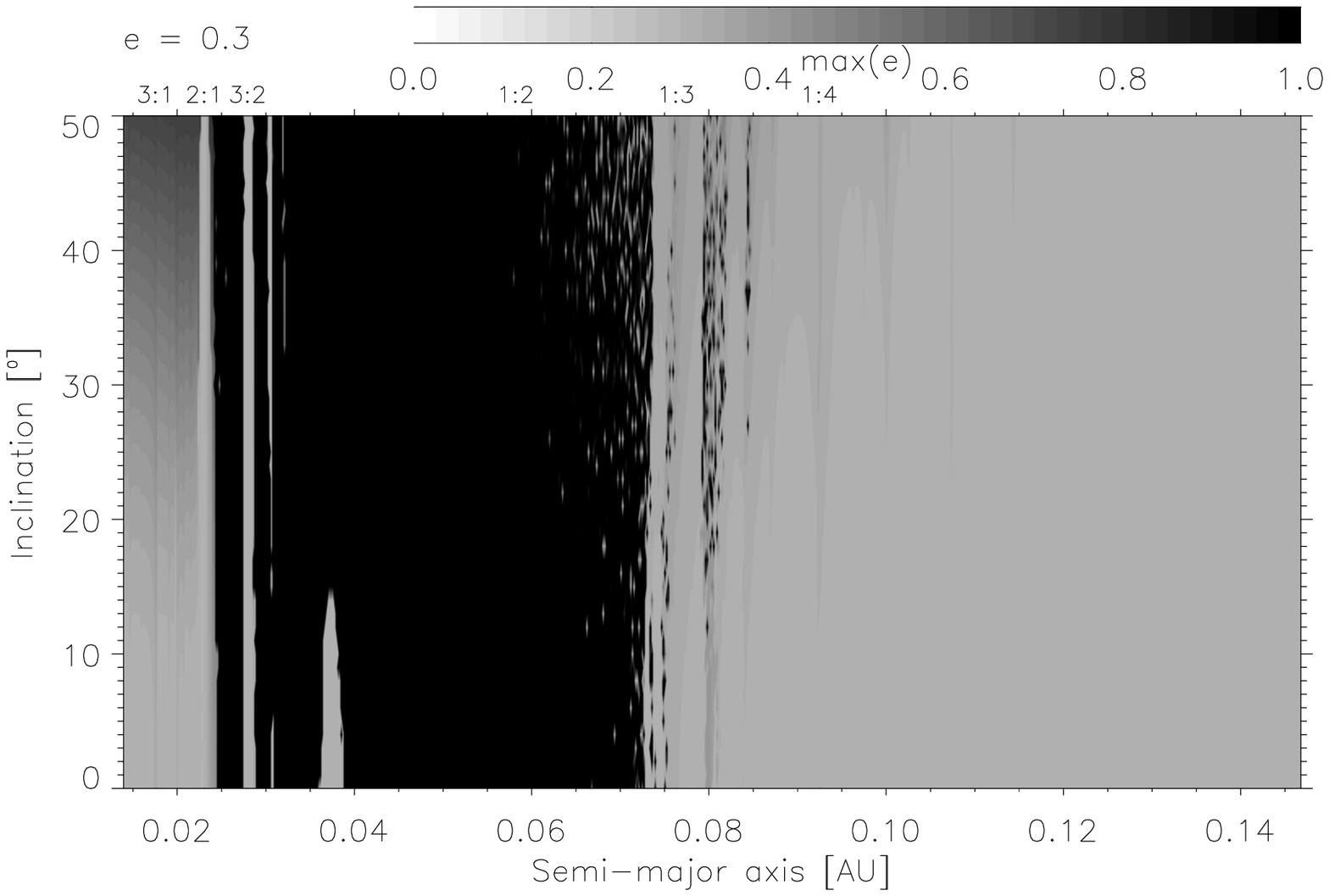}\\
\includegraphics[width=80mm]{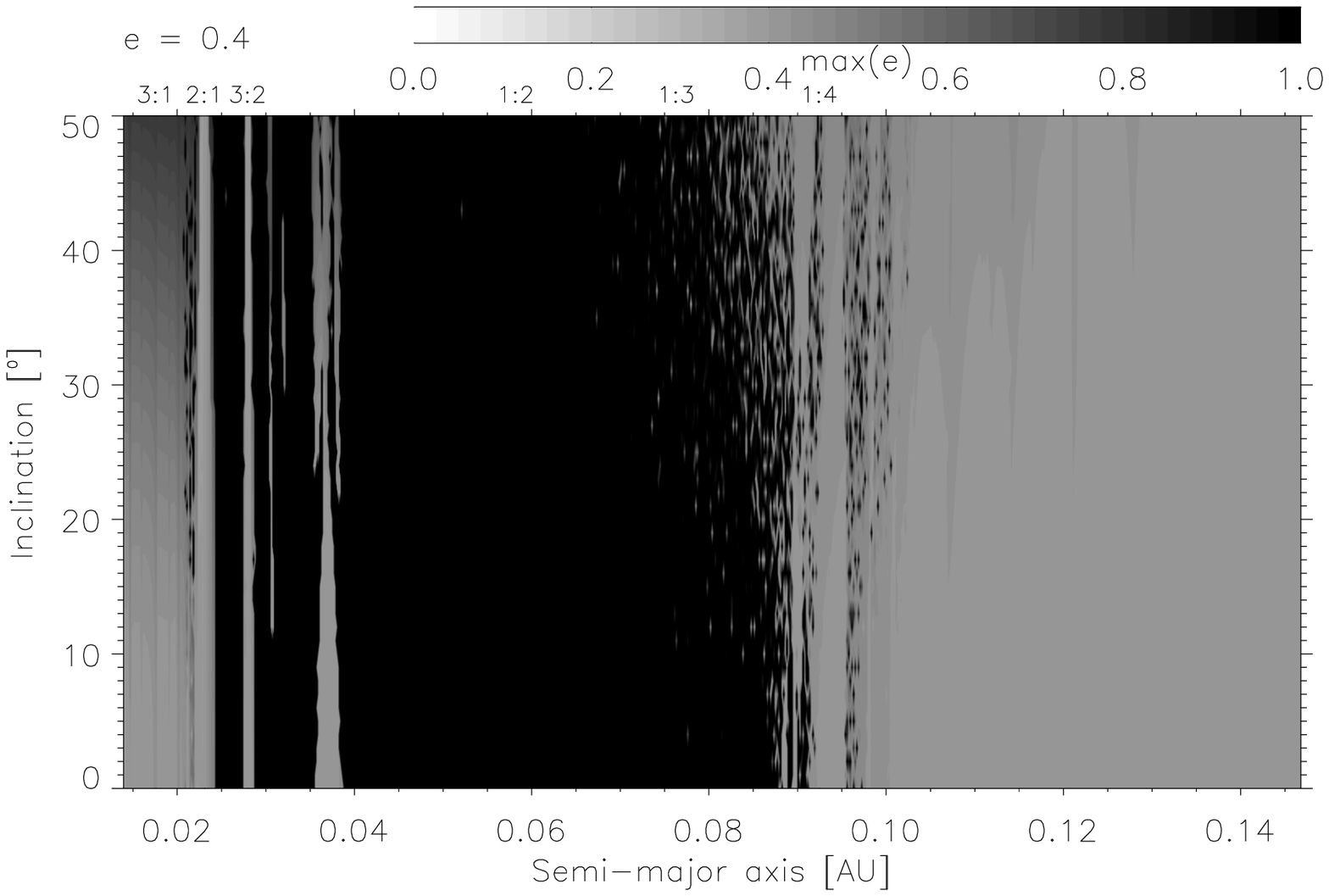}
\includegraphics[width=80mm]{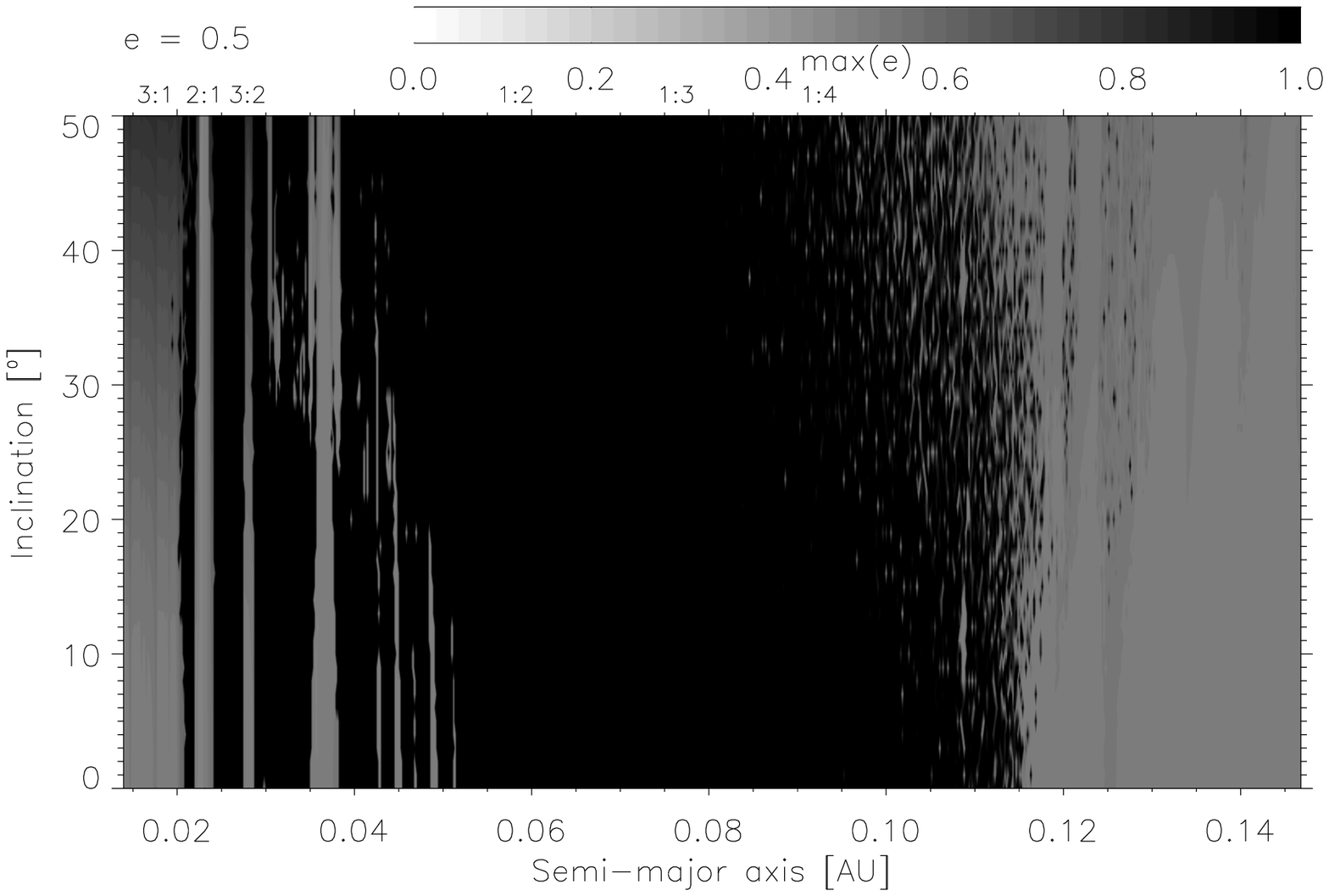}
\end{center}
\caption{Stability plots in the $a-i$ plane for different initial values of $e$ showing the maximum eccentricity (ME). From top left to bottom right: e=0.01, e=0.1, e=0.2, e=0.3, e=0.4, e=0.5.}
\label{label2}
\end{figure*}

The orbital inclination of a planet plays an important role when one wants to observe a transit. We know that - as viewed from the Earth - TrES-2b passes the visible disk of the star. Normally one would expect any additional planets to orbit almost in the same plane as TrES-2b. But from our solar system we know that small objects like asteroids can have substantial orbital inclinations and even the dwarf-planet Pluto has an orbit that is inclined about 17$^\circ$ towards the ecliptic. Knowledge about the possible inclinations that additional planets in the TrES-2 system can have is thus crucial for the search of transits.\\
We thus also investigated the stability regions in the $a-i$ plane; the results can be seen in figure~\ref{label2}. For different initial values of the eccentricity we plotted the maximum eccentricity in the region between 0.014 and 0.147 AU in the semimajor axis and 0$^\circ$ and 50$^\circ$ in the inclination.\\ 
Figure~\ref{label2} shows clearly that the borders of the stability regions behave almost indepently from the initial inclination. For an initial eccentricity of 0.01 the whole region exterior to TrES-2b between 0.05 and 0.147 AU shows ordered motion on circular orbits for all initial values of the inclination (with the exception of the 1:2 and 1:3 resonance). The interior region is also stable up to 0.03 AU although here the orbits become more eccentric for initial inclinations of 30$^\circ$ and larger (with the exception of the 2:1 resonance) where the test particles can maintain almost circular orbits even for larger inclinations.
If the initial eccentricity is increased, the maximum eccentricity increases for all initial values of $a$ and $i$. The border of the exterior stability region moves inward with increasing initial $e$ from 0.05 AU to $\approx$ 0.115 AU for $e=0.5$ (again, at some mean motion resonances the orbits can remain stable). 

\begin{figure*}
\begin{center}
\includegraphics[width=80mm,angle=270]{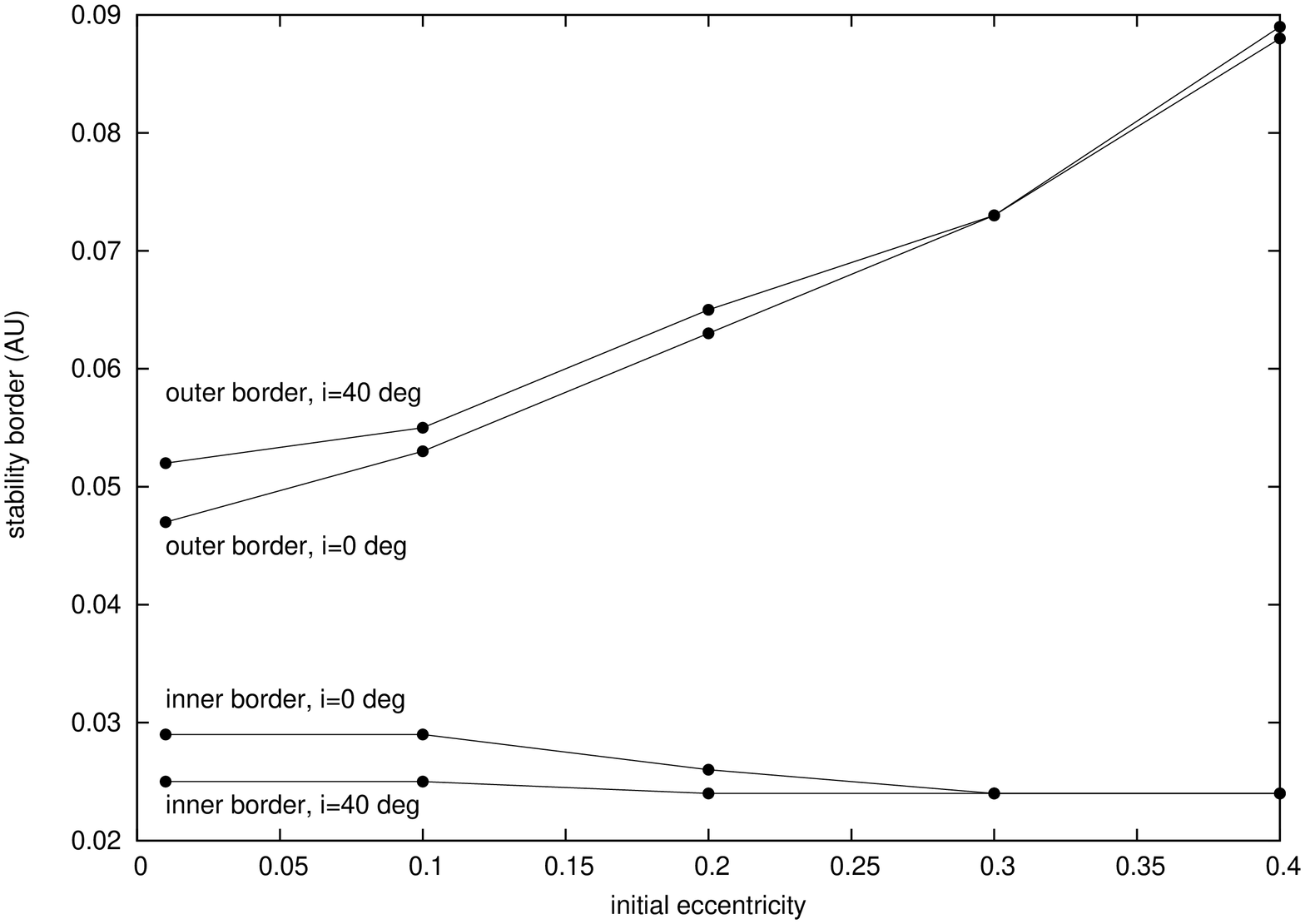}
\end{center}
\caption{Borders of the stability region and their dependence on the initial eccentricity. The borders are shown for i=0$^\circ$ and i=40$^\circ$.}
\label{label2b}
\end{figure*}

Figure~\ref{label2b} shows, how the borders of the stability region depend on the initial eccentricity. The change of the inner border with the initial eccentricity is almost constant whereas the outer border changes linearly (with a slope of approximately $0.1$). The difference between $i=0^\circ$ and $i=40^\circ$ is very small and only significant for small initial eccentricities.

\section{Stability of Resonances}
\label{sec:res}

\begin{figure*}
\begin{center}
\includegraphics[width=110mm]{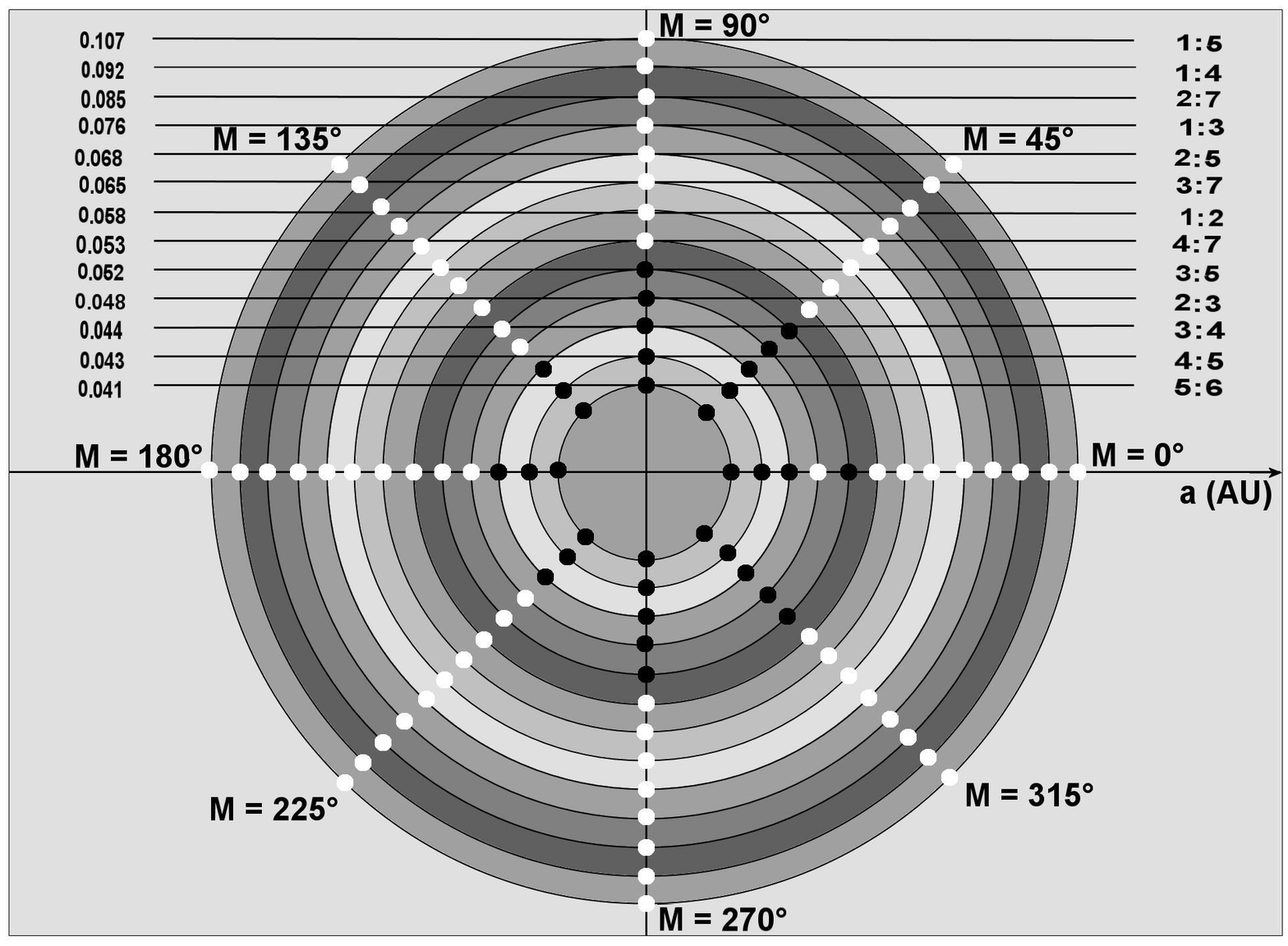}\\
\includegraphics[width=110mm]{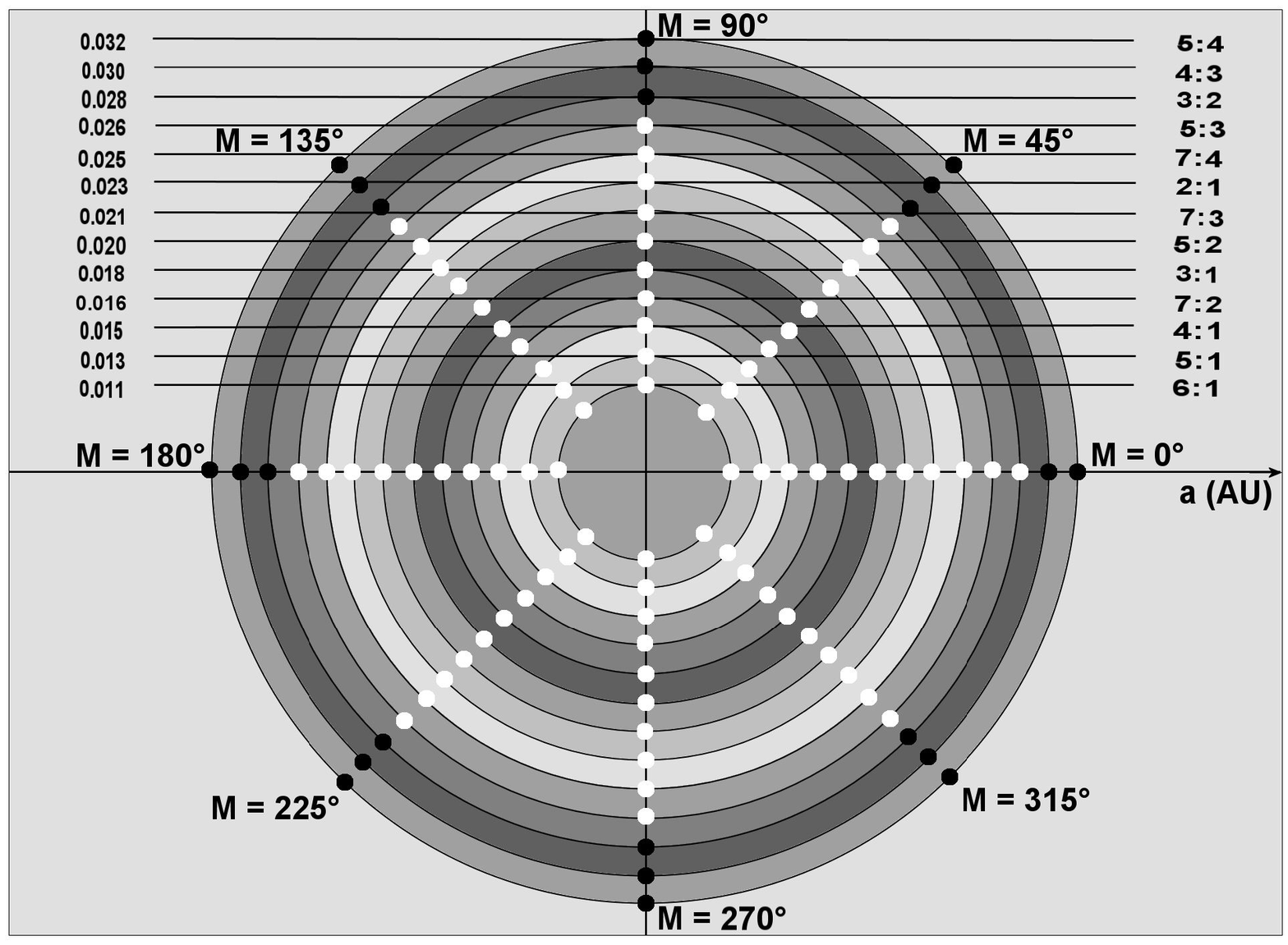}

\end{center}
\caption{Stability of Resonances. The figure shows the stability of the resonances for different initial values of $M$. The upper figure shows the exterior resonances with TrES-2b; the lower figure the interior resonances. White points indicate stable motion; black points show unstable motion.}
\label{label3}
\end{figure*}

Resonances play an important role concerning the possibility of additional planets. On the one hand, a resonant configuration can be more unstable and thus make it possible to exclude larger planets (see e.g.~\cite{miller-ricci08a, miller-ricci08b}). On the other hand, resonances can stabilize the motion of planets under certain conditions and we know about planetary systems that are in a mean-motion resonance (see e.g.~\cite{ferraz05b} and references therein). 

We thus decided to perform a detailed study of the stability inside the mean motion resonances. The stability plots presented in section~\ref{sec:stab} give an overview on the stability properties - but to decide whether a resonance is really stable or not, one has to use special initial conditions. In this simulation the test particles were placed directly inside the resonance. Since the initial mean anomaly can have an important influence on the stability of a resonance (see figure~\ref{label1}) we performed calculations for 8 initial values of $M$ of the test particle between 0$^\circ$ and 315$^\circ$. 

Figure~\ref{label3} shows the results of this study. We investigated all major inner and outer resonances with the known planet. For the outer resonances (figure~\ref{label3} top) we can see, that all resonances between 0.107 AU (1:5) and 0.053 AU (4:7) are stable for all values of $M$. From 0.044 AU (3:4) to 0.041 (5:6) the resonances are unstable for all values of $M$. In between lie the 3:5 and 2:3 resonance where the stability depends on the mean anomaly.\\ 
The inner resonances (figure~\ref{label3} bottom) show a more ordered behavior: from 0.011 AU (6:1) to 0.026 AU (5:3) all resonances are stable; the resonances close to the planet, from 0.028 AU (3:2) to 0.032 AU (5:4) are unstable (with the exception of the 3:2 resonance for $M=0^\circ$).

\section{Conclusions}

We investigated the dynamical stability of the TrES-2 planetary system in order to identify regions where an additional smaller planet could exist. Our results can be summarized as follows:

\begin{itemize}
\item There is broad range of possible orbits for additional planets on almost circular orbits. Such a planet could also exist inside the orbit of TrES-2b. Only the region between approximately $0.03$ and $0.05$ AU is excluded due to the gravitational perturbations of TrES-2b. 
\item Additional planets can also exist on moderately eccentric orbits. Here, the excluded region increases with increasing initial eccentricity. Planets with eccentric orbits ($e>0.4$) can only exist at a distance from the star larger than 0.095 AU.
\item Concerning the stability, a possible inclination of the orbit of an additional planets plays a less important role than its eccentricity. Planets on almost circular orbits and a semimajor axis larger than 0.05 AU can have an orbital inclination up to 50$^\circ$ without becoming dynamically unstable; planets inside the orbit of TrES-2b can be stable up to 35$^\circ$. 
\item Many of the major mean-motion resonances in the system are stable. Only close to the existing planet, they can destabilize the motion of a planet. It is thus possible for planets to exist in a resonant configuration.
\end{itemize}

According to our results, the possibility to detect an additional planet in the TrES-2 system are good - at least from a dynamical point of view. There exists a wide range of regions where such a planet could exist on a stable orbit. Our results also showed that additional planets can have a substantial orbital inclination. If this is the case, a search with the transit technique would fail,
but a detection by the radial velocity technique would be possible.
In this case precise measurements of the transit time of TrES-2b could be used to help identify such a planet.\\
Future works will be dedicated to determine the limits of the masses the additional planets can have. It will then be possible to make much better predictions for the observability of possible transits.



\begin{thebibliography}{}

\bibitem[{{Dvorak} {et~al.}(2003){Dvorak}, {Pilat-Lohinger}, {Funk}, \&
  {Freistetter}}]{dvorak03b}
{Dvorak}, R., {Pilat-Lohinger}, E., {Funk}, B., \& {Freistetter}, F. 2003,
  A\&A, 398, L1

\bibitem[{{Ferraz-Mello} {et~al.}(2005){Ferraz-Mello}, {Michtchenko},
  {Beaug{\'e}}, \& {Callegari}}]{ferraz05b}
{Ferraz-Mello}, S., {Michtchenko}, T.~A., {Beaug{\'e}}, C., \& {Callegari},
  Jr., N. 2005, in LNP Vol. 683: Chaos and Stability in Planetary Systems,
  219--+

\bibitem[{{Ford} \& {Gaudi}(2006)}]{Ford06}
{Ford}, E.~B. \& {Gaudi}, B.~S. 2006, ApJ, 652, L137

\bibitem[{{Ford} \& {Holman}(2007)}]{Ford07}
{Ford}, E.~B. \& {Holman}, M.~J. 2007, ApJ, 664, L51

\bibitem[{{Froeschle}(1984)}]{froeschle84}
{Froeschle}, C. 1984, Celestial Mechanics, 34, 95

\bibitem[{{Hanslmeier} \& {Dvorak}(1984)}]{Hanslmeier84}
{Hanslmeier}, A. \& {Dvorak}, R. 1984, A\&A, 132, 203

\bibitem[{{Lichtenegger}(1984)}]{Lichtenegger84}
{Lichtenegger}, H. 1984, Celestial Mechanics, 34, 357

\bibitem[{{Lohinger} {et~al.}(1993){Lohinger}, {Froeschle}, \&
  {Dvorak}}]{lohinger93a}
{Lohinger}, E., {Froeschle}, C., \& {Dvorak}, R. 1993, Celestial Mechanics and
  Dynamical Astronomy, 56, 315

\bibitem[{{Miller-Ricci} {et~al.}(2008{\natexlab{a}}){Miller-Ricci}, {Rowe},
  {Sasselov}, {Matthews}, {Guenther}, {Kuschnig}, {Moffat}, {Rucinski},
  {Walker}, \& {Weiss}}]{miller-ricci08b}
{Miller-Ricci}, E., {Rowe}, J.~F., {Sasselov}, D., {et~al.} 2008{\natexlab{a}},
  ApJ, 682, 586

\bibitem[{{Miller-Ricci} {et~al.}(2008{\natexlab{b}}){Miller-Ricci}, {Rowe},
  {Sasselov}, {Matthews}, {Kuschnig}, {Croll}, {Guenther}, {Moffat},
  {Rucinski}, {Walker}, \& {Weiss}}]{miller-ricci08a}
{Miller-Ricci}, E., {Rowe}, J.~F., {Sasselov}, D., {et~al.} 2008{\natexlab{b}},
  ApJ, 682, 593

\bibitem[{{Nagy} {et~al.}(2006){Nagy}, {S{\"u}li}, \& {{\'E}rdi}}]{Nagy2006}
{Nagy}, I., {S{\"u}li}, {\'A}., \& {{\'E}rdi}, B. 2006, MNRAS, 370, L19

\bibitem[{{O'Donovan} {et~al.}(2006){O'Donovan}, {Charbonneau}, {Mandushev},
  {Dunham}, {Latham}, {Torres}, {Sozzetti}, {Brown}, {Trauger}, {Belmonte},
  {Rabus}, {Almenara}, {Alonso}, {Deeg}, {Esquerdo}, {Falco}, {Hillenbrand},
  {Roussanova}, {Stefanik}, \& {Winn}}]{Odonovan06}
{O'Donovan}, F.~T., {Charbonneau}, D., {Mandushev}, G., {et~al.} 2006, ApJ,
  651, L61

\bibitem[{{R\"atz} {et~al.}(2008{\natexlab{a}}){R\"atz}, {Mugrauer}, \&
  {Schmidt}}]{Raetz08a}
{R\"atz}, S., {Mugrauer}, M., \& {Schmidt}, T. 2008{\natexlab{a}}, in
  Proceedings of the International Astronomical Union, IAU Symposium, Vol. 253

\bibitem[{{R\"atz} {et~al.}(2008{\natexlab{b}}){R\"atz}, {Mugrauer}, \&
  {Schmidt}}]{Raetz08b}
{R\"atz}, S., {Mugrauer}, M., \& {Schmidt}, T. 2008{\natexlab{b}},
  Astronomische Nachrichten, submitted


\bibitem[{{S{\"u}li} {et~al.}(2005){S{\"u}li}, {Dvorak}, \&
  {Freistetter}}]{suli05}
{S{\"u}li}, {\'A}., {Dvorak}, R., \& {Freistetter}, F. 2005, MNRAS, 363, 241

\bibitem[{{Szebehely}(1967)}]{Szebehely1967}
{Szebehely}, V. 1967, {Theory of orbits. The restricted problem of three
  bodies} (New York: Academic Press)

\bibitem[{{Winn} {et~al.}(2008){Winn}, {Asher Johnson}, {Narita}, {Suto},
  {Turner}, {Fischer}, {Butler}, {Vogt}, {O'Donovan}, \& {Gaudi}}]{Winn08}
{Winn}, J.~N., {Asher Johnson}, J., {Narita}, N., {et~al.} 2008, ArXiv
  e-prints, 0804.2259

\end{thebibliography}
\end{document}